\documentclass[twocolumn,showpacs,preprintnumbers,amsmath,amssymb]{revtex4}
\usepackage{graphics}
\usepackage{graphicx}
\begin{document}

\title{Fidelity susceptibility in the two-dimensional spin-orbit models}
\author{Wen-Long You}
\altaffiliation{Email: wlyou@suda.edu.cn}
\author{Yu-Li Dong}
\affiliation{School of Physical Science and Technology, Soochow
University, Suzhou, Jiangsu 215006, People's Republic of China}
\date{\today}

\begin{abstract}
We study the quantum phase transitions in the two-dimensional
spin-orbit models in terms of fidelity susceptibility and reduced
fidelity susceptibility. An order-to-order phase transition is
identified by fidelity susceptibility in the two-dimensional
Heisenberg XXZ model with Dzyaloshinsky-Moriya interaction on a
square lattice. The finite size scaling of fidelity susceptibility
shows a power-law divergence at criticality, which indicates the
quantum phase transition is of second order. Two distinct types of
quantum phase transitions are witnessed by fidelity susceptibility
in Kitaev-Heisenberg model on a hexagonal lattice. We exploit the
symmetry of two-dimensional quantum compass model, and obtain a
simple analytic expression of reduced fidelity susceptibility.
Compared with the derivative of ground-state energy, the fidelity
susceptibility is a bit more sensitive to phase transition. The
violation of power-law behavior for the scaling of reduced fidelity
susceptibility at criticality suggests that the quantum phase
transition belongs to a first-order transition. We conclude that
fidelity susceptibility and reduced fidelity susceptibility show
great advantage to characterize diverse quantum phase transitions in
spin-orbit models.
\end{abstract}


\maketitle
\section{Introduction}
Spin-spin interactions have been intensively studied in quantum
magnets and Mott insulators in the last decades
\cite{Zaanen-PRL-55-418}. The effect of the orbital degree of
freedom has received much attention since the discovery of a variety
of novel physical phenomena and a diversity of new phases in
transition metal oxides (TMOs) \cite{Science288.462,NatMater6.927}.
In particular, under an octahedral environment, the $d$ orbital
degeneracy of transition metal ions is partially lifted, and the
remaining orbital degrees of freedom can be generally described by
localized $S=1/2$ pseudospins. A typical effect induced by such a
symmetry degradation of orbital degeneracy is the presence of
bond-selective pseudospin interaction. It is because the spatial
orientations of the orbits lead to anisotropic overlaps between
neighboring ions. Consequently, the interactions among different
bonds are intrinsically frustrated. To understand the orbital degree
of freedom, the so-called quantum compass model (QCM) has been given
rise to intensive research
\cite{SovPhysJETP.37.725,PhysRevLett.93.207201,PhysRevB.71.024505,PhysRevB.72.024448,PhysRevB.71.195120,PhysRevB.75.144401,PhysRevB.75.134415,PhysRevB.78.064402,PhysRevLett.102.077203,PhysRevB.80.014405,PhysRevB.78.184406,PhysRevB.79.104429}.
In QCM, the pseudospin operators are coupled in such a way as to
mimic the built-in competition between the orbital orders in
different directions \cite{Science288.462,NatMater6.927}. The
frustration leads to macroscopic degeneracy in the classical ground
state \cite{PhysRevLett.93.207201}, and even highly degenerate
quantum ground state \cite{PhysRevB.75.134415,PhysRevB.78.184406}.
Interestingly, the two-dimensional (2D) QCM has become a prototype
to generate topologically protected qubits
\cite{PhysRevB.71.024505,PhysRevLett.99.020503}.

Besides, considering the spin-orbit coupling in TMOs is inevitable
and intriguing. For example, octahedra tilt may give rise to
effective Dzyaloshinsky-Moriya interaction (DMI)
\cite{PhysRevLett.102.017205}. Comparing those that only possess
spin-spin coupling, spin-orbit models appear to be more intricate.
The interplays between spin and orbital degree of freedom host a
variety of different phases. For instance, the orbital exchange in a
honeycomb lattice induces orbital ordering
\cite{Wu-PRL-2006,PhysRevB.75.195118} and topological order
\cite{Kitaev}.

In this paper, we concentrate on portraying the quantum phase
transitions (QPTs) in 2D spin-orbit interaction Hamiltonians. As we
know, a QPT identifies any point of nonanalyticity in the
ground-state (GS) energy of an infinite lattice system
\cite{Sachdev}. Conventionally, local order parameters are needed to
detect the nonanalyticity in the GS properties as the system varies
across the quantum critical point (QCP). However, the knowledge of
the local order parameter is not easy to retrieve from a general
many-body system, especially for QPTs beyond the framework of the
Landau-Ginzburg symmetry breaking paradigm
\cite{Kitaev,Feng-PRL-98-087204}. Recently, quantum fidelity, also
referred to as the GS fidelity, sparked great interest among the
community to use it as a probe for the QCP
\cite{HTQuan2006,Pzanardi2006}. The fidelity defines the overlap
between two neighboring ground states of a quantum Hamiltonian in
the parameter space, i.e.,
\begin{eqnarray}
F (\lambda,
\delta\lambda)=|\langle\Psi_0(\lambda)|\Psi_0(\lambda+\delta\lambda)\rangle|,
\label{eq:fidedifintion}
\end{eqnarray}
where $\vert \Psi_0(\lambda) \rangle$ is the GS wave function of a
many-body Hamiltonian $H(\lambda)$=$H_0$ + $\lambda H_I$, $\lambda$
is the external driving parameter, and $\delta \lambda$ is a tiny
variation of the external parameter. Though borrowed from the
quantum information theory, fidelity has been proved to be a useful
and powerful tool to detect and characterize QPTs in condensed
matter physics \cite{IntJModPhys.24.4371}. In order to remove the
artificial variation of external parameters, one of the authors and
collaborators in Ref. \cite{PhysRevE.76.022101} introduced the
concept of fidelity susceptibility (FS), which is the leading-order
term of fidelity with respect to the external driving parameter,
\begin{eqnarray}
\chi_{\textrm F}\equiv \lim_{\delta\lambda\rightarrow 0}\frac{-2\ln
F}{\delta\lambda^2}.
\end{eqnarray}
FS elucidates the rate of change of fidelity under an infinitesimal
variation of the driving parameter. There exists an intrinsic
relation between the FS and the derivatives of GS energy as
following:
\begin{eqnarray}
\chi_{\textrm F}(\lambda) &=& \sum_{n \neq 0} \frac{\vert \langle
\Psi_n(\lambda)\vert H_I \vert \Psi_0(\lambda)\rangle \vert^2
}{\left[E_0(\lambda) - E_n(\lambda)\right]^2}, \label{expansechi}\\
\frac{\partial ^2 E_0(\lambda)}{\partial \lambda^2} &=& \sum_{n \neq
0} \frac{\vert \langle \Psi_n(\lambda)\vert H_I \vert
\Psi_0(\lambda)\rangle \vert^2 }{E_0(\lambda) - E_n(\lambda)}.
\label{expanseE0}
\end{eqnarray}
Here the eigenstates $\vert \Psi_n(\lambda)\rangle$ satisfy
$H(\lambda) \vert \Psi_n(\lambda)\rangle = E_n(\lambda)\vert
\Psi_n(\lambda)\rangle$. The apparent similarity between Eq.
(\ref{expansechi}) and Eq. (\ref{expanseE0}) arouses similar
critical behavior around the critical point.

 In addition, the reduced fidelity and
its susceptibility were also suggested in the studies of QPTs
\cite{Paunkovic,Jian-PRE-78-051126,Ho-PRA-78-062302,You-PLA}. The
reduced fidelity concerns the similarity of a local region of the
system with respect to the driving parameter. Despite the locality,
the reduce fidelity fidelity (RFS) encodes the fingerprint of QPTs
of the whole system, and is even more sensitive than its global
counterparts \cite{Eriksson-PRA-2009,Zhi-PRA-2010}.

For a second-order QPT, around the critical point, the correlation
length $\xi$ diverges as $(\lambda -\lambda_c)^{-\nu}$, while the
gap in the excitation spectrum vanishes as $(\lambda
-\lambda_c)^{z\nu}$, where $\lambda_c$ is the critical point in the
thermodynamic limit, $\nu$ and $z$ are the critical exponents. At
the critical point, i.e., $\xi = \infty$, the only length scale is
the system size $L$. We can account for the divergence at the QPT by
formulating a finite-size scaling (FSS) theory. Universal
information could be decoded from the scaling behavior of FS
\cite{Kwok-PRB-2008,Gu-PRB-2008,Yu-PRE-2009}. FS increases as the
system size grows, and the summation in Eq.(\ref{expansechi})
contributes to an extensive scaling of $\chi_{\textrm{F}}$ in the
off-critical region. Therefore, FS per site $\chi_{\textrm{F}}/N$
appears to be a well-defined value, where $N=L^d$ is the number of
sites and $d$ stands for system dimensionality. Instead, FS exhibits
stronger dependence on system size across the critical point than in
non-critical region, showing that a singularity emerges in the
summation of Eq.(\ref{expansechi}). This implies an abrupt change in
the ground state of the system at the QCP in the thermodynamic
limit. Following standard arguments in scaling analysis
\cite{Continentino}, one obtains that FS per site scales as
\cite{Schwandt-PRL-2009,Grandi,arXiv1104.4104}
\begin{eqnarray}
\chi_{\textrm{F}}/N \sim L^{2/\nu-d}. \label{chiscaling}
\end{eqnarray}
Due to the arbitrariness of relevance of the driving Hamiltonian
$H_I$ under the renormalization group transformation,
$\chi_{\textrm{F}}/N$ could be (i) superextensive if $\nu d <2 $
\cite{Schwandt-PRL-2009}, (ii) extensive if $\nu d =2$
\cite{Garnerone,Jacobson-PRB-2009}, (iii) subextensive if $\nu d >2$
\cite{Venuti-PRL-99-095701}. From Eq. (\ref{expansechi}) we can
deduce that the ground state of the system should be gapless if
$\chi_{\textrm{F}}$ is superextensive, but not vise versa. For
finite size system, the position of a divergence peak defines a
pseudocritical point $\lambda_L^*$ as the precursor of a QPT, and it
approaches the critical point $\lambda_c$ as $L \to \infty$. For
sufficiently relevant perturbations on large size system, i.e., $\nu
d<2$, the leading term in expansion of pseudocritical point obeys
such scaling behavior as \cite{Zanardi-PRA-2008}
\begin{eqnarray}
|\lambda_L^* - \lambda_c | \sim L ^{-1/\nu}.\label{lambdascaling}
\end{eqnarray}
Thus, the behavior of $\chi_{\textrm{F}}$ on finite systems in the
vicinity of a second-order QCP can be estimated as
\cite{Schwandt-PRB-2010,arXiv1106.4078,arXiv1106.4031}
\begin{eqnarray}
 \chi_{\textrm{F}}/N \approx C L^{2/\nu-d} f(|\lambda-\lambda_c| L^{1/\nu}), \label{collapsescaling}
 \end{eqnarray}
where $f$ is an unknown regular scaling function, and $C$ is a
constant independent of $\lambda$ and $L$.

However, many important QPTs fall in the category of first-order
QPTs, in which the first derivative of GS energy exhibits
discontinuity at critical point. Different from second-order QPT,
there is no characteristic correlation length $\xi$ in first-order
QPT, and in general the FSS will violate the scaling relations of
second-order QPTs \cite{Binder-RPP-1987}. For a typical first-order
QPT, two competing ground states $\vert \Psi_{<}(\lambda) \rangle$
and $\vert \Psi_{>}(\lambda) \rangle$ are degenerate at critical
point $\lambda_c$ in the thermodynamic limit, and they become
energetically favorable on one side of $\lambda_c$ such as $\vert
\Psi_{<}(\lambda) \rangle =\vert \Psi_0(\lambda<\lambda_c) \rangle$
and $\vert \Psi_{>}(\lambda) \rangle =\vert
\Psi_0(\lambda>\lambda_c) \rangle$ respectively. The level crossings
in the thermodynamic limit usually turn into avoided level crossings
for finite system, and degeneracy at critical point in general is
lifted, opening an energy gap $\Delta_g$. In the low-energy
subspaces spanned by $\vert \Psi_{<}(\lambda_c) \rangle$ and $\vert
\Psi_{>}(\lambda_c) \rangle$, the diagonal matrix elements coincide
$\langle \Psi_{<}(\lambda_c) \vert H \vert \Psi_{<}(\lambda_c)
\rangle $=$\langle \Psi_{>}(\lambda_c) \vert H \vert
\Psi_{>}(\lambda_c) \rangle$, but the off-diagonal matrix elements
$\langle \Psi_{<}(\lambda_c) \vert H \vert \Psi_{>}(\lambda_c)
\rangle $= $(\langle \Psi_{>}(\lambda_c) \vert H \vert
\Psi_{<}(\lambda_c) \rangle)^* $ induce an avoided level crossing
with
\begin{eqnarray}
\Delta_g = 2 \vert \langle \Psi_{<}(\lambda_c) \vert H \vert
\Psi_{>}(\lambda_c) \rangle \vert. \label{gap}
\end{eqnarray}
For a Hamiltonian with local interactions (e.g., nearest neighbors),
\begin{eqnarray}
\langle \Psi_{<}(\lambda_c) \vert H \vert \Psi_{>}(\lambda_c)
\rangle = \sum_{m,n \in {\cal S}} c_m^*c_n H_{mn},
\label{matrxielement}
\end{eqnarray}
where $H_{mn}=\langle \varphi_m \vert H \vert \varphi_n \rangle$,
$\vert \Psi_{<}(\lambda_c) \rangle = \sum_{m \in S} c_m
\vert\varphi_m\rangle$, $\vert \Psi_{>}(\lambda_c) \rangle = \sum_{n
\in {\cal S}} c_n \vert \varphi_n \rangle$, $\vert
\varphi_{m}\rangle$ ($\vert \varphi_{n}\rangle$) form a complete set
of basis vectors in Hilbert space ${\cal S}$, and $c_{m}$ ($c_{n}$)
are the corresponding coefficients. The dimension of ${\cal S}$ is
of ${\cal D}^N$ (${\cal D}$ is the degree of freedom of Hamiltonian
constituent, e.g., ${\cal D}=2$ for spin-1/2 Hamiltonian), and
$c_{m}$ ($c_{n}$) are of order ${\cal D}^{-N/2}$. The finite Hamming
distance between $\vert\varphi_{m}\rangle$ and
$\vert\varphi_{n}\rangle$ gives limited number (roughly speaking,
${\cal O}(N)$) of nonzero $H_{mn}$. Hence, the off-diagonal matrix
elements scale exponentially $\langle \Psi_{<}(\lambda_c) \vert H
\vert \Psi_{>}(\lambda_c) \rangle \sim {\cal O}(N{\cal D}^{-N})$,
and consequently $\Delta_g \sim  {\cal O}(N{\cal D}^{-N})$
\cite{Schutzhold,Jorg}. If any of the ground states are degenerate
protected by symmetries, $H$ should be written in the subspace of
all the degenerate states, and the dimension of the matrix will be
larger than $2 \time 2$. Considering Eq. (\ref{expansechi}), the
exponentially closing gap gives a dominant contribution, manifesting
that the FS $\chi_F$ should carry the signature of an exponential
divergence at critical point. Moreover, from another point of view,
due to their macroscopic distinguishability, the overlap between
states $\vert \langle \Psi_{<}(\lambda) \vert \Psi_{>}(\lambda)
\rangle \vert $ should be exponentially small with system size $N$
at criticality \cite{Schutzhold}. In general, FS per site scales
exponentially, i.e.,
\begin{eqnarray}
\chi_F/N \sim g(N) e^{- \mu N },\label{scalingof1storder}
\end{eqnarray}
where $\mu$ is a size independent constant, and $g(N)$ is a
polynomial function of $N$, which is a correction to the exponential
term.

In this work, we use FS and RFS to incarnate the critical phenomena
in 2D spin-orbit models. We show that the FS and RFS manifest
themselves as extreme points at the QCPs. In other words, FS and RFS
are quite sensitive detectors of the QPTs in spin-orbit models. This
article is organized as follows: In Sec. \ref{General_Hamiltonian},
a general Hamiltonian is given. We investigate QPTs in the
spin-orbit model on a square lattice in Sec. \ref{XXZmodelwithD}.
The phase diagram and the FSS are studied through FS. Next, in Sec.
\ref{2DKitaev-Heisenbergmodel}, we study QPTs in Kitaev-Heisenberg
model on a 2D honeycomb lattice by FS and the second derivative of
the the GS energy. Both approaches identify three distinct phases,
and FS behaves more sensitively. After that, we take advantage of
the RFS to locate the QCP of 2D compass model in Sec.
\ref{2DCompassModel}, and show that the exponentially divergent
peaks of FS and RFS imply that the first order QPT. A brief summary
is presented in Sec. \ref{Summary}.

\section{General Hamiltonian}
\label{General_Hamiltonian} Motivated by a broad interest in Mott
insulators, we construct the general structure of model Hamiltonian
on a given $\gamma$ bond connecting two nearest-neighbor (NN) sites
$\mathbf{i}$ and $\mathbf{j}$,
\begin{eqnarray}
{\cal H_{\mathbf{ij}}^\gamma}  =  J_{\gamma} S_\mathbf{i}^{\gamma}
S_\mathbf{j}^{\gamma} + J \vec{S}_\mathbf{i} \cdot
\vec{S}_\mathbf{j} + \vec{D} \cdot ( \vec{S}_\mathbf{i} \times
\vec{S}_\mathbf{j} ).\label{GeneralHamiltonian}
\end{eqnarray}
The first term in Hamiltonian (\ref{GeneralHamiltonian}) describes
bond-selective interaction, where the orbital exchange constant
$J_{\gamma}=4 t_{\gamma}^2/U$ is derived from multi-orbital Hubbard
Hamiltonian consisting of the local on-site repulsion $U$ and the
hopping integral $t_{\gamma}$. The second term corresponds to the
isotropic Heisenberg coupling, and Dzyaloshinsky-Moriya anisotropy
$\vec{D}$ in the third term comes from the lattice distortions. The
Hamiltonian (\ref{GeneralHamiltonian}) extrapolates from the
Heisenberg model to QCM depending on the bonds geometry. In this
paper, we devote our study to 2D lattices, either square or
hexagonal lattice. Several real systems motivate the investigation
of these kinds of lattices, such as SrTiO$_3$
\cite{Schrimer-JPC-17-1321}, Sr$_2$IrO$_4$
\cite{Cao-RPB-57-R11039,Moon-PRB-74-113104,Kim-PRL-101-076402}. The
flexibility of parameters induces a rich variety of phases and
various fascinating physical phenomena.

\section{Fidelity Susceptibility in the Spin-Orbit model}
\label{XXZmodelwithD} In strong spin-orbit coupling materials, an
XXZ model with DMI was suggested,

\begin{eqnarray}
 {\cal H}_{\textrm{DM}}(\Delta, D) &=& J \sum_{<\mathbf{ij}>} {S}_\mathbf{i}^x
 {S}_\mathbf{j}^x +  {S}_\mathbf{i}^y {S}_\mathbf{j}^y + \Delta S_\mathbf{i}^z
S_\mathbf{j}^z
  \nonumber \\ &+&  \vec{D}\cdot  ( \vec{S}_\mathbf{i} \times
\vec{S}_\mathbf{j}),\label{SO-Hamiltonian}
\end{eqnarray}
where $\Delta$ is the anisotropic parameter and $D$ is the strength
of DMI. A general boundary condition can be written as
$S_{N+1}^{\pm}=pS_0^{\pm}$, where $S_{\mathbf{i}}^{\pm}=
{S}_\mathbf{i}^x \pm i {S}_\mathbf{i}^y$, $p = 0$ corresponds to
open boundary condition (OBC) and $p = 1$ to the periodic boundary
condition (PBC). The Hamiltonian (\ref{SO-Hamiltonian}) was proposed
to describe the layered compound Sr$_2$IrO$_4$
\cite{PhysRevLett.102.017205}. The spin canting is induced by
lattice distortion of corner-shared IrO$_6$ octahedra. The DMI was
introduced originally to explain the presence of weak ferromagnetism
in antiferromagnetic (AFM) materials
\cite{Dzyaloshinsky-JPCS-4-241,Moriya-PR-4-288}, such as
$\alpha$-Fe$_2$O$_3$, MnCO$_3$ and CoCO$_3$, since such
antisymmetric interaction could produce small spin cantings.
Recently, the influence of DMI has become very important in
elucidating many interesting properties of different systems, e.g.,
ferroelectric polarization in multiferroic materials
\cite{Sergienko-PRB-2006}, exchange bias effects in perovskites
\cite{Dong-PRL-2009}, asymmetric spin-wave dispersion in double
layer Fe \cite{Zakeri-PRL-2010,Michael-PRB-2010}, and noncollinear
magnetism in FePt alloy films \cite{Honolka-PRL-2009}. For
simplicity, we assume the DMI fluctuation is acting on the $x$-$y$
plane, and the vector $\vec{D}$ is imposed along the $z$ direction, i.e., $\vec{D}%
=D\vec{z}$.

The Hamiltonian (\ref{SO-Hamiltonian}) reduces to the anisotropic
Heisenberg XXZ model when the rotations of IrO$_6$ octahedra are
absent, i.e., $D$=0. When $\Delta \gg 1$, one-dimensional (1D)
spin-1/2 XXZ chain has long range order and gapped domain-wall
excitations. On the other hand, in the XX limit, i.e., $\Delta$ = 0,
the system is equivalent to a chain of non-interacting fermionic
model, which becomes gapless in the thermodynamic limit. The QPT
takes place at the isotropic point $\Delta_c$ = 1, which is a
Berezinskii-Kosterlitz-Thouless (BKT) transition point. BKT phase
transition belongs to an infinite-order phase transition, and the
ground-state energy and all of its derivatives with respect to
$\Delta$ are continuous at the critical point. However, the FS
succeeds in detecting the nonanalyticities of the ground state
across BKT transition \cite{Yang-PRB-2007}. For instance, the BKT
transition from gapless Tomonaga-Luttinger liquid to gapped Ising
phase in 1D XXZ model is detected by the divergence of the FS using
the density-matrix-renormalization-group (DMRG) technique
\cite{Wang-PRA-2010}. In addition, the BKT transition from spin
fluid to dimerized phase in the $J_1$-$J_2$ model
\cite{Chen-PRA-2008} and the superfluid-insulator transition in the
Bose-Hubbard model at integer filling \cite{Gu-PRB-2008} are also
able to be signaled by FS.

Different from the 1D case, 2D XXZ model exhibits a second-order QPT
at the isotropic point $\Delta_c$=1 in the thermodynamic limit,
where the first excited energy levels cross \cite{Tian-PRB-2003}.
For $\Delta \gg 1$, the Ising term in the Hamiltonian dominates and
the ground state is an AFM phase along the $z$ direction. For
$\Delta \ll 1$, the first two terms in the Hamiltonian dominate and
the ground state is also an AFM phase, but in the $x$-$y$ plane. It
is well known that long-range orders are present in both phases
\cite{Kennedy-PRL-1988}. FS can serve as a sensitive detector of the
critical point in the 2D XXZ model on a relatively small square
lattice \cite{Yu-PRE-2009}.

Actually, the DMI does not change the universality class of XXZ
model. The DMI can be eliminated from the Hamiltonian
(\ref{SO-Hamiltonian}) by a spin axes rotation
\cite{Perk,Aristov-PRB-2000,Alcaraz-JSP-1990}, and after rotation an
unitarily equivalent form is given by
\begin{eqnarray}
 {\cal H}_{\textrm{DM}}(\Delta, D) \sim \frac{1}{\cos \phi}{\cal
 H}_{\textrm{XXZ}}(\widetilde{\Delta}),
\end{eqnarray}
where $\tan \phi = D$, $\widetilde{\Delta}=\Delta \cos \phi$ and
${\cal H}_{\textrm{XXZ}}(\widetilde{\Delta})=J \sum_{<\mathbf{ij}>}
{S}_\mathbf{i}^x
 {S}_\mathbf{j}^x +  {S}_\mathbf{i}^y
 {S}_\mathbf{j}^y + \widetilde{\Delta} S_\mathbf{i}^z
S_\mathbf{j}^z $ with boundary condition $ S_{N+1}^{\pm}=pe^{\mp
i(N\phi)}S_0^{\pm}$. Note that the mapping becomes exactly
equivalent only in the thermodynamic limit and for open boundary
condition. In the thermodynamic limit, the boundary condition does
not affect the critical behavior and consequently the ${\cal
H}_{\textrm{DM}}(\Delta, D)$ will have the same critical properties
as the ${\cal H}_{\textrm{XXZ}} (\widetilde{\Delta})$. Hence, QCP at
$\Delta_c$=1 in ${\cal H}_{\textrm{XXZ}}$ becomes a critical line
$\Delta_c = \sqrt{1+D^2}$ in ${\cal H}_{\textrm{DM}}$.

\begin{figure}[tbp]
\includegraphics[width=9cm]{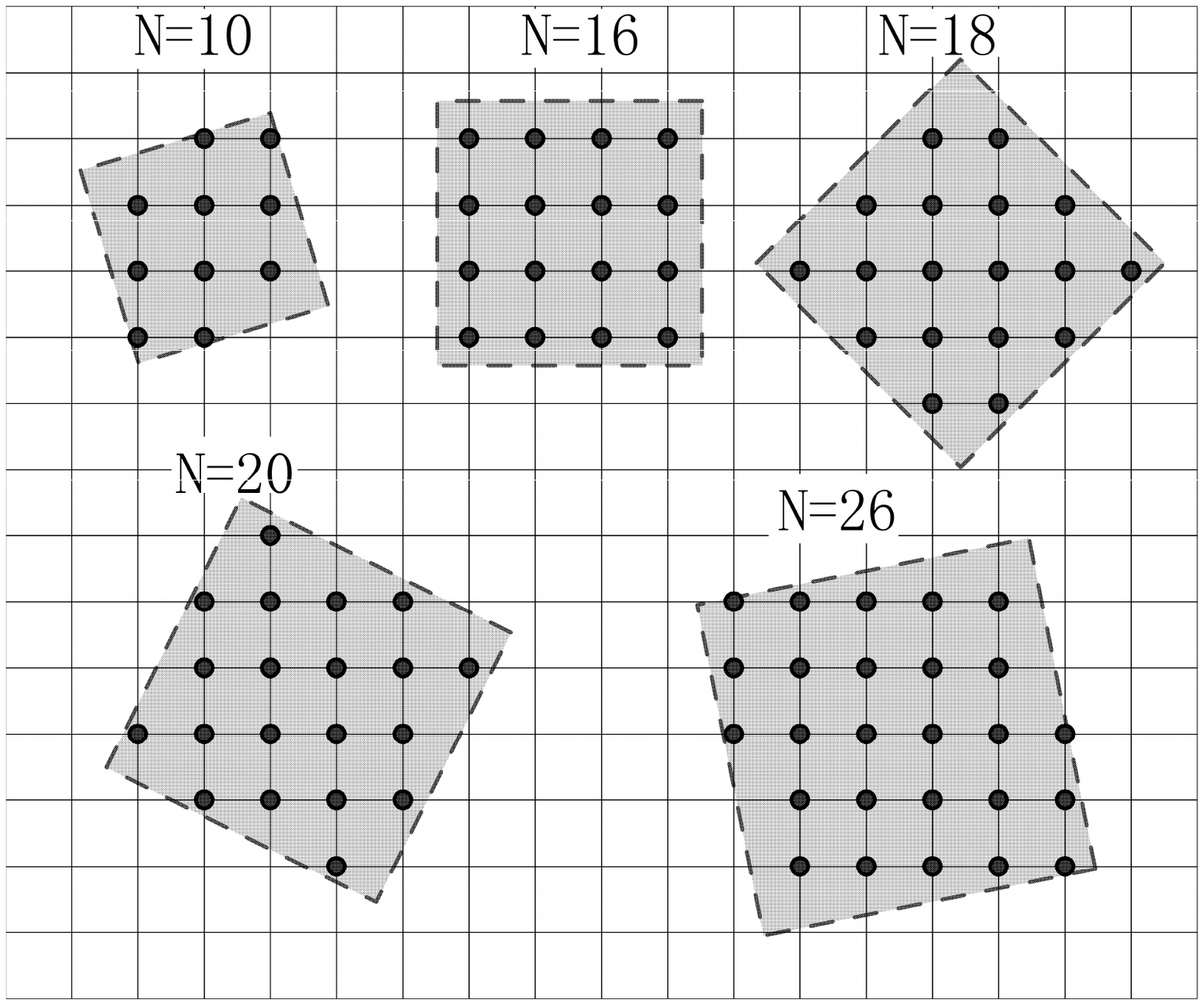} %
\caption{(Color online) Two-dimensional square structures for
different system sizes N=10,16,18,20,26.} \label{2Dsquarelattice}
\end{figure}

\begin{figure}[tbp]
\includegraphics[width=8cm]{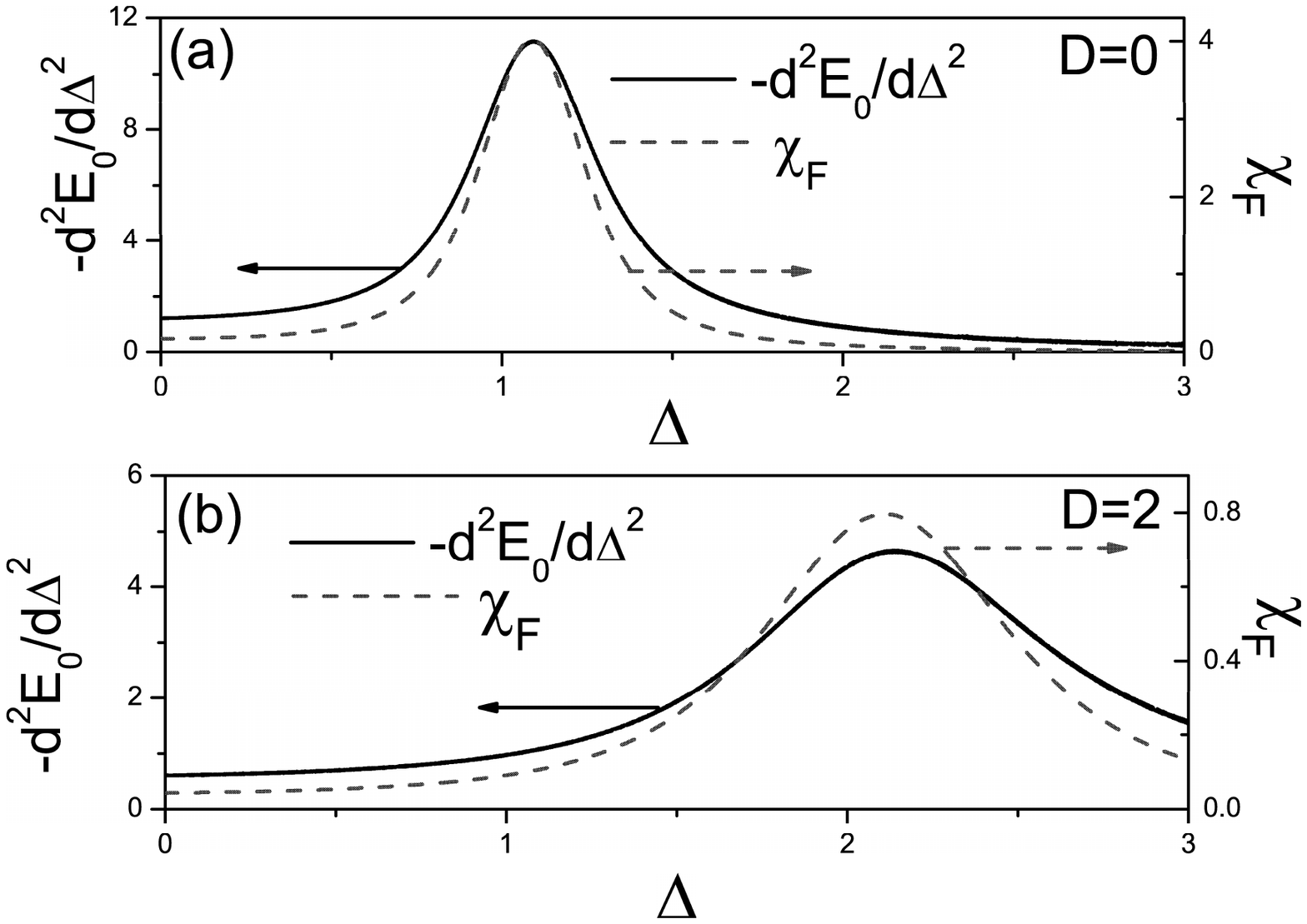} %
\caption{(Color online) The second derivative of ground-state energy
(solid line) and fidelity susceptibility (dash line) of $N=20$
square lattice as a function of $\Delta$ for (a) $D=0$ and (b)
$D=2$. } \label{N20-E-FS}
\end{figure}

FS and the second derivative of GS energy of $N=20$ square lattice
(see Fig. \ref{2Dsquarelattice}) as a function of anisotropy $\Delta$
for $D=0$ and $D=2$ are demonstrated in Fig. \ref{N20-E-FS}. As is
sketched, both approaches display broad peaks at an identical
$\Delta$. However, the peak of FS is slightly narrower than that of
second derivative of GS energy. The locations of peaks are specified
as pseudocritical points. The heights of local maxima decrease as
$D$ increases.

\begin{figure}[tbp]
\includegraphics[width=9cm]{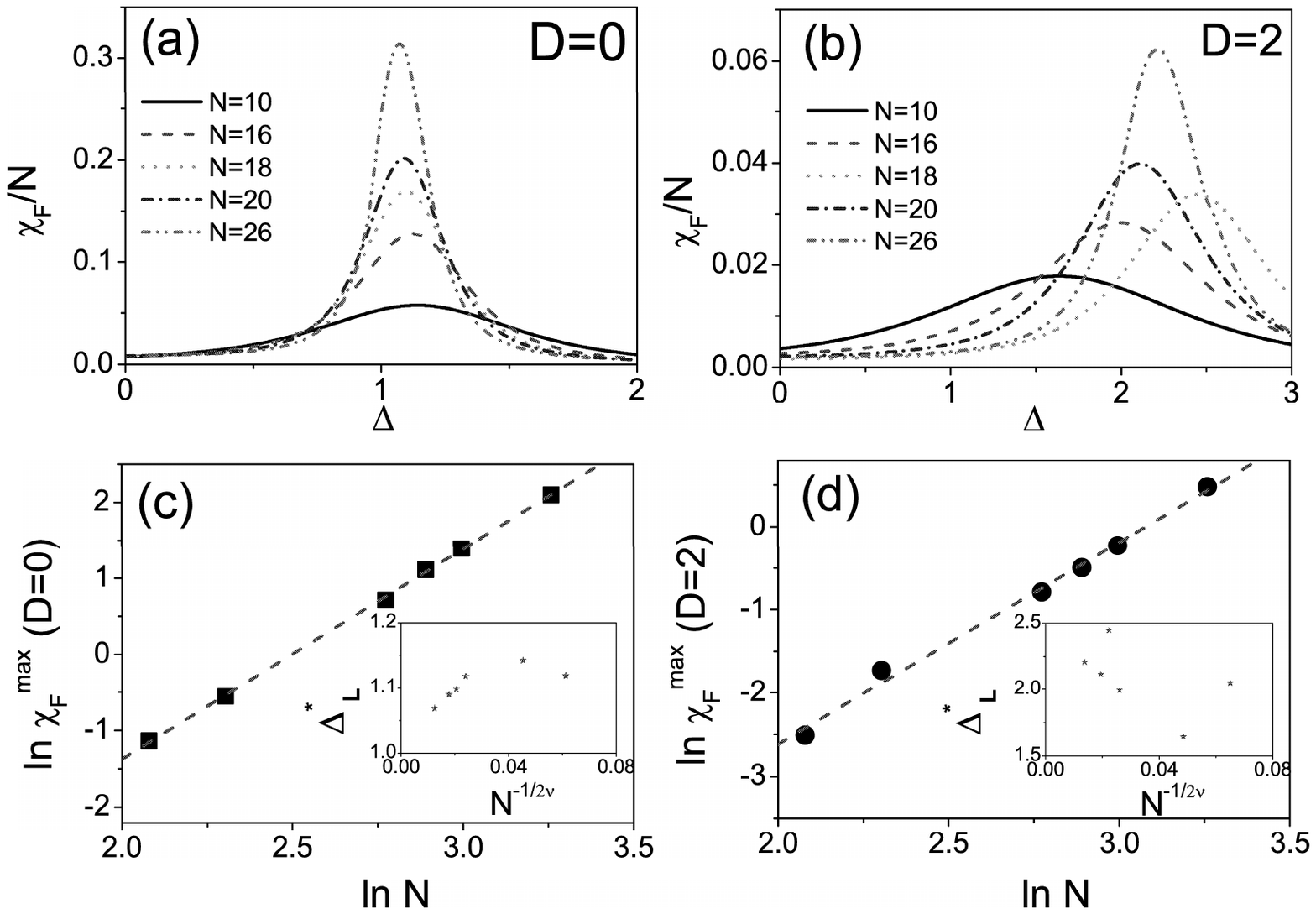} %
\caption{(Color online) Top: the fidelity susceptibility per site
$\chi_{\textrm{F}}/N$ as a function of anisotropy parameter $\Delta$
for various size $N$= 10, 16, 18, 20, 26 with (a) $D=0$ and (b)
$D=2$. Bottom: finite size scaling analysis for the height and
location of the peaks. The logarithm of maximum fidelity
susceptibilities $\chi_{\textrm{F}}^{\textrm{max}}$ for (c) $D$=0
and (d) $D$=2, respectively, are plotted as a function of $\ln N$.
The dash lines are least square straight line fits with $ 1/\nu
\approx 2.68(6)$ for (c) $D$=0 and $ 1/\nu \approx 2.62(8)$ for (d)
$D$=2. The insets in panels (c) and (d) show pseudocritical points
$\Delta_L^*$ for various sizes $N$ correspondingly. }
\label{fig:XF-Delta-N}
\end{figure}

Next, to check whether these extreme points can be regarded as QCPs,
we should perform FSS analysis, as is illustrated in Fig.
\ref{fig:XF-Delta-N}. In Figs. \ref{fig:XF-Delta-N}(a) and
\ref{fig:XF-Delta-N}(b), we plot FS per site as a function of
$\Delta$ for different lattice size $N$=10, 16, 18, 20 and 26. As is
clearly shown that $\chi_{\textrm{F}}$ becomes more pronounced for
increasing $N$. $\chi_{\textrm{F}}$ is extensive in the off-critical
region, while superextensive at the pseudocritical point. From the
scaling relation Eq. (\ref{chiscaling}), a linear dependence for the
maximum fidelity susceptibilities $\chi_{\textrm{F}}^{\textrm{max}}$
on $L^{2/\nu }$ is expected with effective length $L=\sqrt{N}$. This
is confirmed by the results shown in Figs. \ref{fig:XF-Delta-N}(c)
and \ref{fig:XF-Delta-N}(d), in which we plot the logarithm of
maximum fidelity susceptibilities $\chi_{\textrm{F}}^{\textrm{max}}$
at pseudocritical points verse a function of $\ln N$. For large $N$,
$\ln \chi_{\textrm{F}}^{\textrm{max}}$ scales linearly with $\ln N$,
and the slope is interpreted as the inverse of critical exponent of
the correlation length $1/\nu$ from Eq. (\ref{chiscaling}). By
applying linear regression to the raw data obtained from XXZ model
on various square lattices, we derive $ 1/\nu \approx 2.68(6)$ for
$D=0$, while $ 1/\nu \approx 2.62(8)$ for $D=2$. The values got from
the measurements of  $D=0$ and $D=2$ are consistent with each other
up to two digits. The validation of Eq. (\ref{chiscaling}) implies
that the QPT in 2D XXZ with DMI manifests itself as a clear sign of
a second-order QPT.

From Figs. \ref{fig:XF-Delta-N}(a) and \ref{fig:XF-Delta-N}(b), the
positions of cusps seemingly converge toward the critical points.
However, these system sizes seem insufficiently to estimate the
critical exponents from Eqs. (\ref{lambdascaling}) and
(\ref{collapsescaling}), and the corrections to scaling relations
are not negligible. In this case, the shift of the location of the
pseudocritical point from real critical point should be replaced
with $ |\Delta_L^* - \Delta_c | \approx c_1 L ^{-1/\nu} + c_2 L
^{-2/\nu} + \ldots$, where $c_1$ and $c_2$ are the coefficients
\cite{Zanardi-PRA-2008}. As is shown in insets of Figs.
\ref{fig:XF-Delta-N}(c) and \ref{fig:XF-Delta-N}(d), we notice that
the data points corresponding to the small system sizes clearly
deviate from the linear fit obtained for the points for the two
largest $N$. As $N \to \infty$, the pseudocritical points
$\Delta_L^*$ approach the critical points $\Delta_c$. The linear
fits from results of $N=20$ and $N=26$ (see Fig.
\ref{2Dsquarelattice}) yield the estimates the $\Delta_c$ are $1.02$
and $2.47$, respectively. To get more precise critical exponents,
more elaborate schemes are needed, such as large-scale quantum Monte
Carlo simulations \cite{Schwandt-PRL-2009,Schwandt-PRB-2010}.

\begin{figure}[tbp]
\includegraphics[width=8cm]{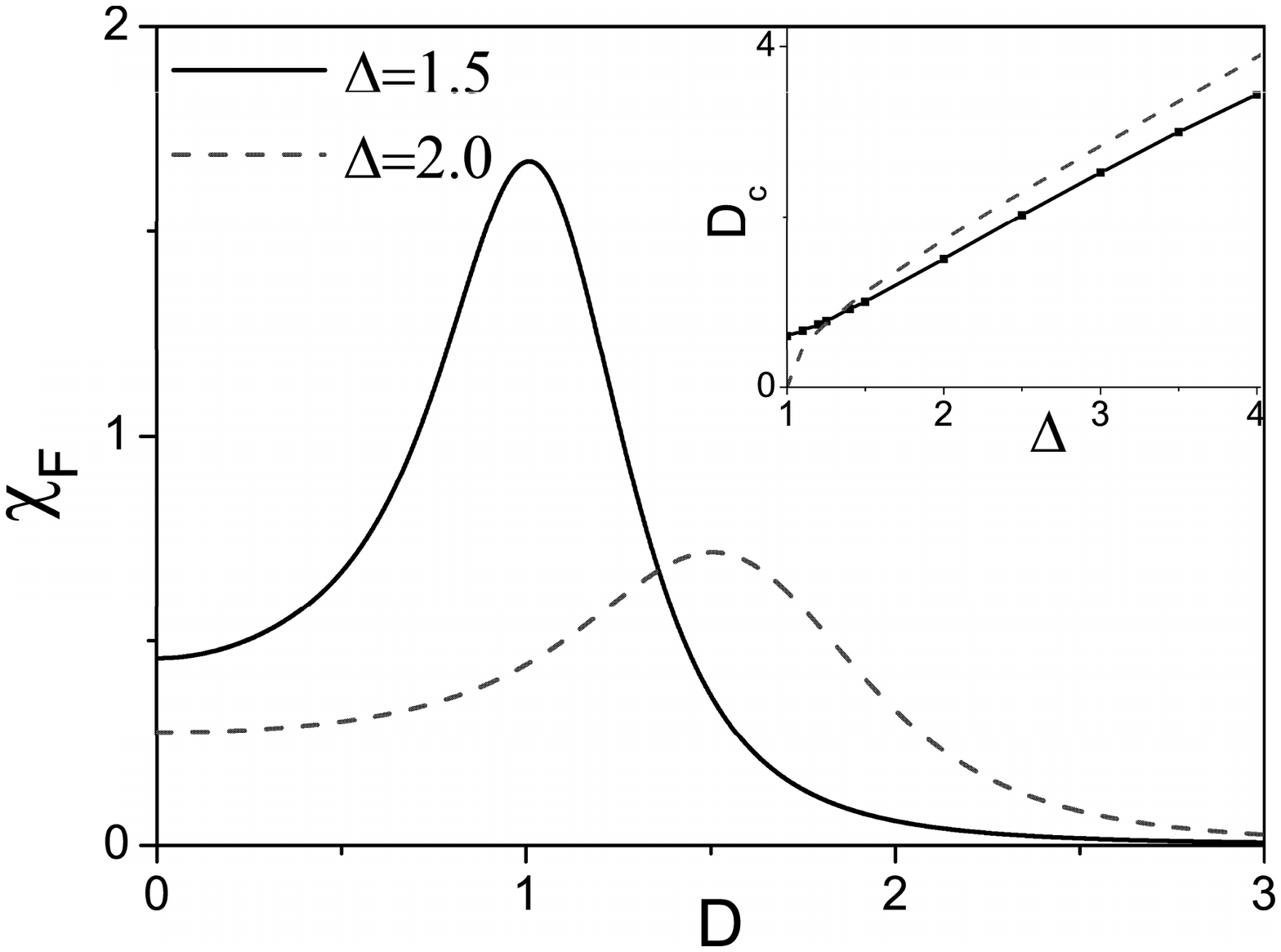} %
\caption{(Color online) The fidelity susceptibility
$\chi_{\textrm{F}}$ as a function of $D$ for $\Delta=1.5$ and
$\Delta=2.0$. Inset shows the pseudocritical line retrieved from the
peaks of fidelity susceptibility for $N=20$ (solid dotted line) and
the critical line $D_c = \sqrt{\Delta^2-1}$ in thermodynamic limit
(dash line).} \label{N20-Delta-D}
\end{figure}

\begin{figure}[tbp]
\includegraphics[width=10cm]{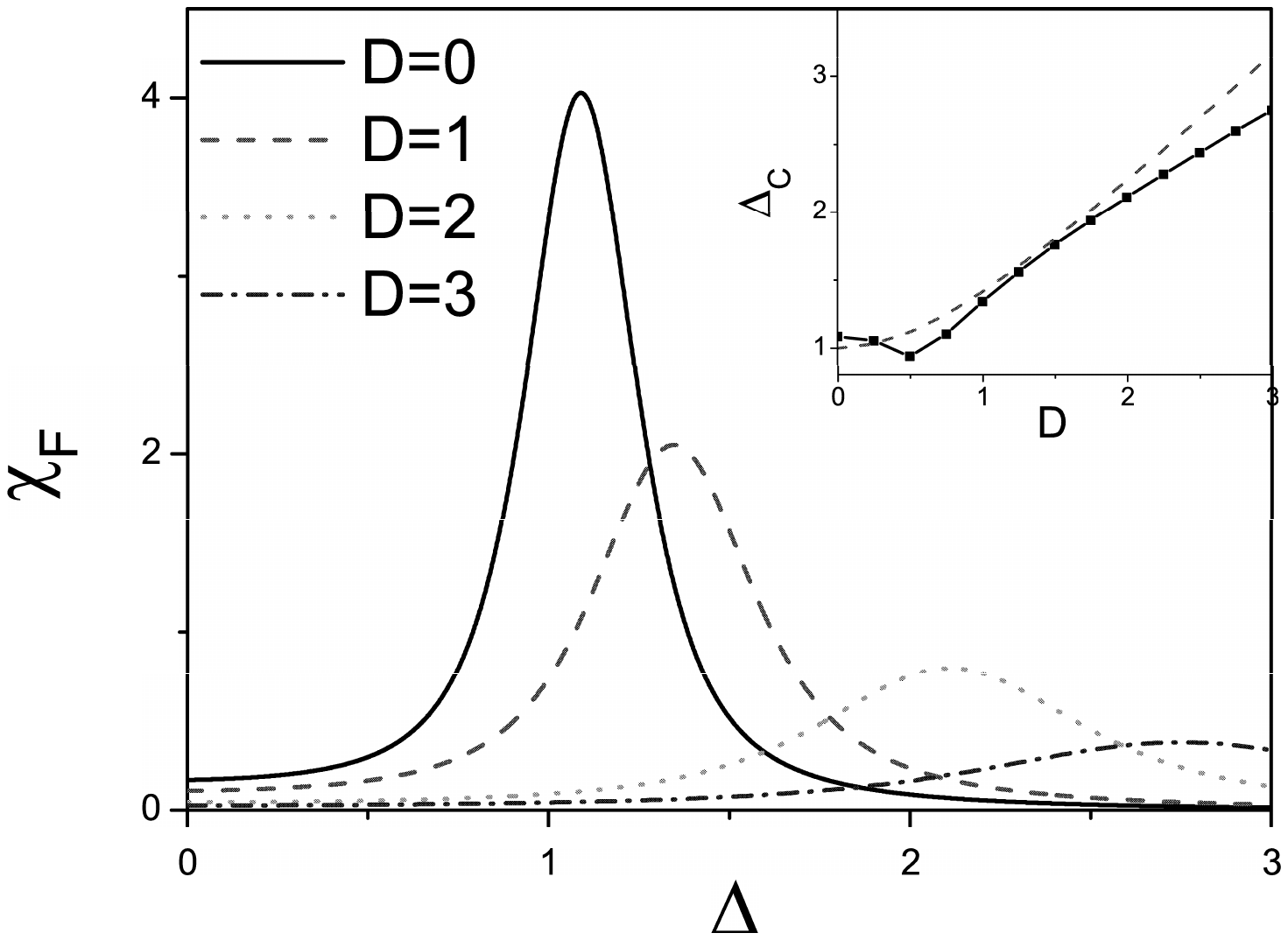} %
\caption{(Color online) The fidelity susceptibility
$\chi_{\textrm{F}}$ in the ground state of the 2D 20-site XXZ model
in terms of anisotropy for different values of DMI. Inset shows the
pseudo critical line by the extraction from the maximum points of
the fidelity susceptibility (solid dotted line), and the critical
line $\Delta_c = \sqrt{1+D^2}$ in thermodynamic limit (dash line).}
\label{XF-Delta-N20}
\end{figure}

To proceed, we calculate the FS on $N$=20 square lattice for
different sets of parameters $\Delta$ and $D$. The FS as a function
of $D$ for $\Delta=1.5$ and $\Delta=2.0$ are depicted in Fig.
\ref{N20-Delta-D}, and the FS as a function of $\Delta$ for $D$=0,
1, 2, and 3 are plotted in Fig. \ref{XF-Delta-N20}. A boost of
$\Delta$ and $D$ suppress the FS. By retrieving the critical points
from the locations of FS peaks, we are able to draw the
corresponding phase diagrams. The pseudocritical lines in the $(D,
\Delta)$ plane and the $(\Delta, D)$ plane are shown in the insets
of Fig. \ref{N20-Delta-D} and Fig. \ref{XF-Delta-N20}. We can notice
that pseudocritical line for $N=20$ (solid dotted line) are
qualitatively similar to the critical line in the thermodynamic
limit (dash line).

\section{Fidelity susceptibility in the 2D Kitaev-Heisenberg model}
\label{2DKitaev-Heisenbergmodel} 
\begin{figure}[tbp]
\includegraphics[width=8cm,bb=0 177 587 672]{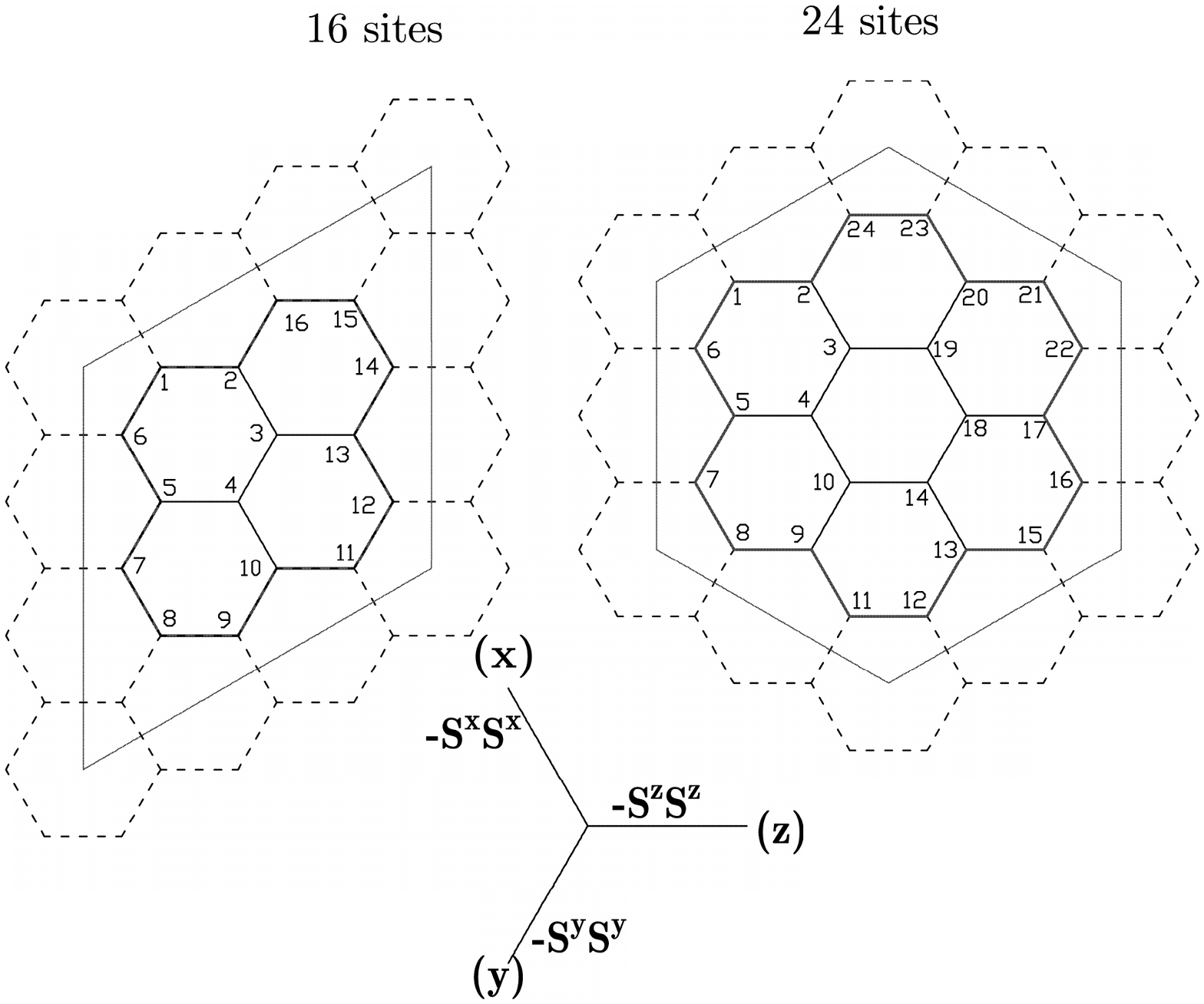}
\caption{(Color online) Two-dimensional structures for system sizes
$N$=16 and 24, which can be placed on a hexagonal lattice with
periodic boundary conditions.} \label{fig:216revised5-Model}
\end{figure}

Since Kitaev introduced a spin 1/2
quantum lattice model with Abelian and non-Abelian topological
phases, finding a physical realization of Kitaev model has triggered
a tremendous amount of interest \cite{PhysRevLett.105.027204}. It is
proposed that in iridium oxides A$_2$IrO$_3$, the strong spin-orbit
coupling may lead to the desired anisotropy of the spin interaction.
The Ir$^{4+}$ ions in iridium oxides A$_2$IrO$_3$ can be effectively
illustrated as spin half on a honeycomb lattice, where three
distinct types of NN bonds are referred to as $\gamma=$ ($0^\circ$,
$120^\circ$, $240^\circ$) bonds. To describe the competition between
direct exchange and superexchange mechanisms, a so-called
Kitaev-Heisenberg model was dedicated, i.e.,
\begin{eqnarray}
%{\cal H}^{\gamma}_{ij}
{\cal H}_{\textrm{KH}}=\sum_{<{\mathbf i}{\mathbf j}>||\gamma} -2
\alpha S_{\mathbf i}^{\gamma} S_{\mathbf j}^{\gamma} + (1-\alpha)
\mathbf{S}_{\mathbf i} \cdot \mathbf{S}_{\mathbf j}.
\label{Kitaev-Heisenberg-Model}
\end{eqnarray}
The Hamiltonian (\ref{Kitaev-Heisenberg-Model}) has been
parameterized, which encompasses a few well-known models. For
$\alpha=0$, it reduces to the Heisenberg model on a hexagonal
lattice, while it becomes exactly solvable ferromagnetic Kitaev
model at $\alpha=1$. Another solvable point corresponds to
$\alpha=1/2$, where it is unitarily equivalent to ferromagnetic
Heisenberg model. The ground state of Hamiltonian
(\ref{Kitaev-Heisenberg-Model}) evolves from 2D N\'{e}el AFM state
($\alpha=0$) to stripy antiferromagnetism ($\alpha=1/2$), and to
spin liquid ($\alpha=1$) as $\alpha$ increases. We perform exact
diagonalization (ED) on two geometries with $N$=16 and $N$=24 (see
Fig. \ref{fig:216revised5-Model}) to calculate the GS energy $E_0$
and fidelity susceptibility $\chi_{\textrm{F}}$.

\begin{figure}[tbp]
\includegraphics[width=8cm]{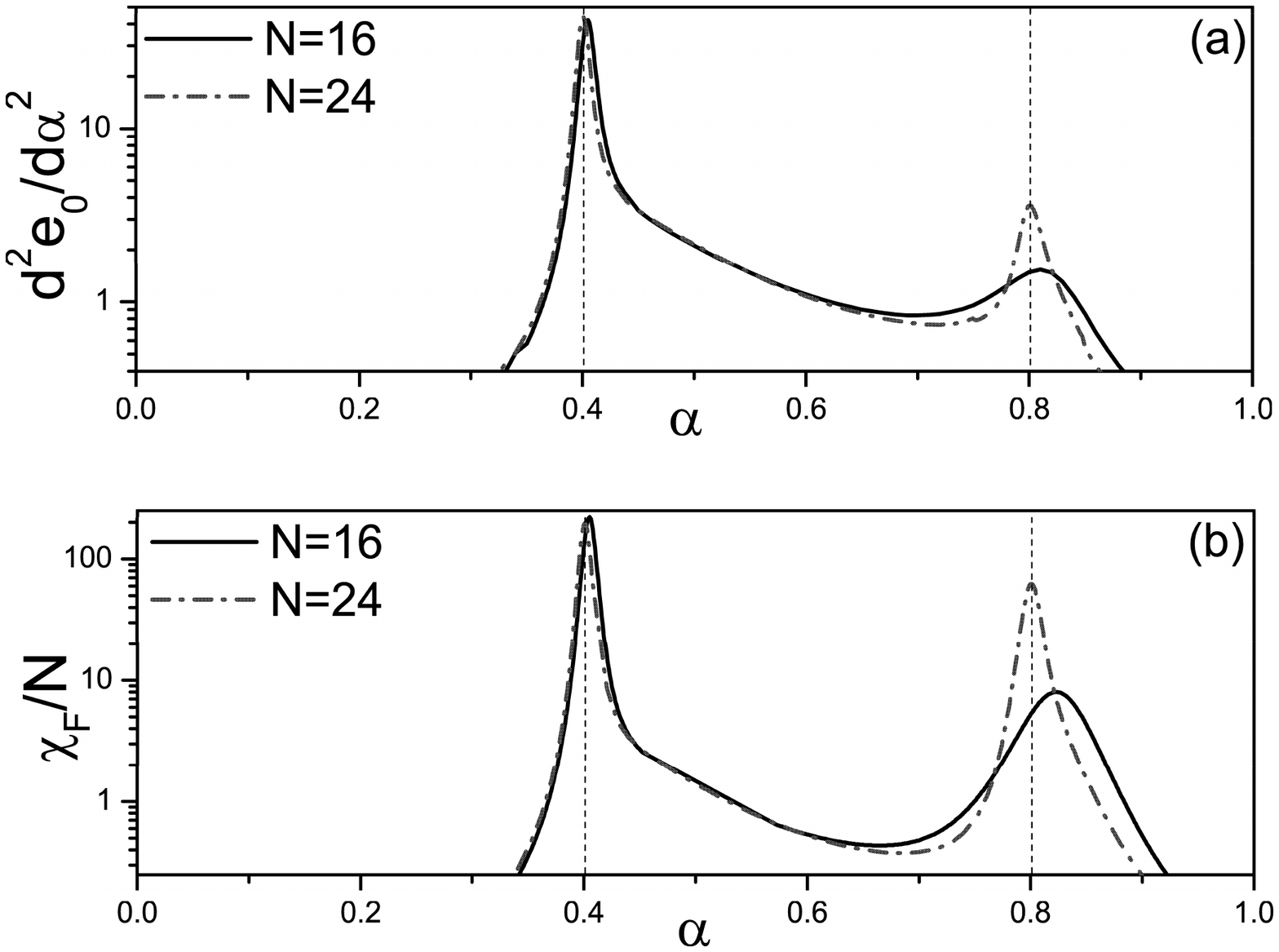}
\caption{(Color online) (a) The second derivative of energy density
$e_0$ verse $\alpha$ for $N=16$ and $N=24$. (b) Fidelity
susceptibility per site $\chi_{\textrm{F}}/N$ as a function of
$\alpha$. } \label{fig:FS-E-16-24}
\end{figure}

The second derivative of energy density $e_0 \equiv - E_0 /N$ and FS
per site $\chi_{\textrm{F}}/N$ as a function of coupling strength
$\alpha$ are obtained by Lanczos calculation. As is illustrated in
Fig. \ref{fig:FS-E-16-24}, two anomalies emerge when $\alpha$
increases from $0$ to $1$, indicating the system undergoes two phase
transitions. The characterization of QPTs by FS is compatible with
the second derivative of energy density. We find that they play
similar roles in identifying the QPTs \cite{Chen-PRA-2008}. However,
the former demonstrates more pronounced peaks than the latter
(notice the $y$-axis is of a logarithmic scale). QPT from
N\'{e}el-ordered to stripe-ordered phase takes place at $\alpha
\approx 0.4$, and this order-to-order phase transition seems
insensitive to system size $N$ for these two configurations. On the
other hand, the second order-to-disorder phase transition at $\alpha
\approx 0.8$ appears to be sensitive to $N$. The numerical
calculations point out that the QPT around $\alpha = 0.4$ is
first-order, while there is no consensus on the character of the QPT
around $\alpha=0.8$ \cite{HCJiang,Fabien}. However, it is difficult
to carry on ED on larger cluster for the hexagonal geometry and
substantiate the scaling behavior of 2D Kitaev-Heisenberg model
unless sophisticated techniques are used
\cite{Schwandt-PRL-2009,Schwandt-PRB-2010}.

\section{Reduced Fidelity Susceptibility in The 2D Compass model}
\label{2DCompassModel} In recent years, the 2D AFM QCM has attracted
considerable attention due to its interdisciplinary character. On
one hand, it plays an important role in describing orbital
interactions in TMOs. On the other hand, 2D QCM is equivalent to
Xu-Moore model \cite{Xu2004} and toric code model in a transverse
field \cite{Vidal2009}, which could possibly be used to generate
protected qubits realized by Josephson-coupled $p\pm ip$
superconducting arrays \cite{PhysRevB.71.024505}.

QCM is defined on a $N=L \times L$ square lattice with PBC by the
Hamiltonian
\begin{equation}
%\begin{split}
 {\cal H}_{\textrm{QCM}}=J_x\sum_{\langle \mathbf{i,j}\rangle ||\mathbf{e_x}}S_{\mathbf{i}}^x S_{\mathbf{j}}^x+J_z\sum_{\langle\mathbf{i,j}\rangle ||\mathbf{e_z}}S_{\mathbf{i}}^z
  S_{\mathbf{j}}^z, \label{CompassModel}
\end{equation}
where $\mathbf{e_x,e_z}$ are unit vectors along the $x$ and $z$
directions, and $J_x$ ($J_z$) is the coupling in the $x$ ($z$)
direction. $S_{\mathbf{i}}^\alpha$ ($\alpha$=$x$, $y$, $z$) are the
pseudospin operators of lattice site $\mathbf{i}$ obeying $
[S_{\mathbf{i}}^\alpha,S_{\mathbf{j}}^\beta]=i\epsilon_{\alpha \beta
\gamma}S_{\mathbf{i}}^\gamma
  \delta_{\mathbf{i,j}}$. There exists a first-order phase transition at
the self-dual point, i.e., $J_x= J_z$, and it is widely believed
that the phase transition belongs to first-order
\cite{PhysRevB.72.024448,PhysRevB.75.144401,PhysRevLett.102.077203,Vidal2009}.

In the case of $L$ being even, this model is equivalent to the
ferromagnetic QCM by rotating the pseudospin operators at one
sublattice by an angle $\pi$ about $\hat{S}^y$-axis. The Hamiltonian
(\ref{CompassModel}) enjoys such commutation relation with the
parity operators $\hat{P}_j$ and $\hat{Q}_i$, which are defined as
\begin{equation}
\hat{P}_j  = \prod_{i=1}^L \left(2\hat{S}^z_{i,j}\right) =
\prod_{i=1}^L \hat{\sigma}^z_{i,j}, \>\>\>\>\>\>\> \hat{Q}_i  =
\prod_{j=1}^L \left(2\hat{S}^x_{i,j}\right) = \prod_{j=1}^L
\hat{\sigma}^x_{i,j}, \label{Symmetry Operators}
\end{equation}
where indices $i$ and $j$ are the $x$- and $z$-component of lattice
site $\bf i$. However, column parity operator $\hat{P}_j$ does not
commute but anticommutes with row parity operator $\hat{Q}_i$. In
such circumstance, the Hilbert space can then be decomposed into
subspace $V(\{p_j\})$, where $p_j $ is the eigenvalue of $\hat{P}_j
$ and is specified in each subspace. Since $\hat{P}_j^2=1 $, $p_j $
can be either 1 or -1. The lowest energy of the Hamiltonian in the
subspace $V(\{p_j\})$ is nondegenerate. In Ref.
\cite{Brzezicki-PRB-2010}, the authors unraveled the hidden dimer
order therein. The symmetries in these systems induce large
degeneracies in their energy spectra, and make their numerical
simulation tricky. Recently, the ground state of ${\cal
H}_{\textrm{QCM}}$ is proven to reside in the most homogeneous
subspaces $V(\{p_j=1\})$ and  $V(\{p_j=-1\})$, in other words, they
are two-fold degenerate \cite{You-JPA-2010}.

\begin{figure}[tbp]
\includegraphics[width=8cm]{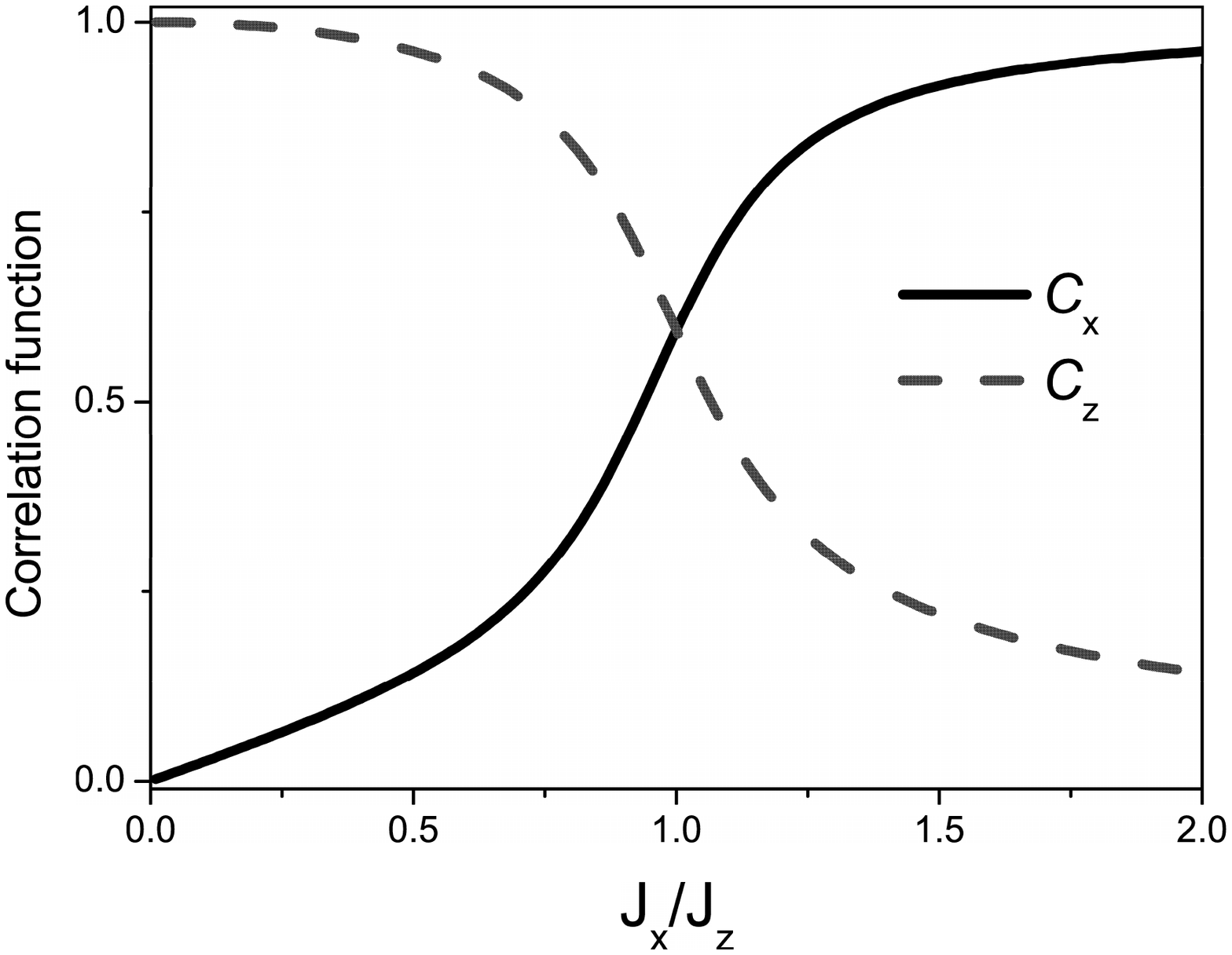}
\caption{(Color online) The nearest-neighbor correlation functions
$\mathcal{C}_x$ and $\mathcal{C}_z$ with respect to $J_x/J_z$ for
$N=16$ square lattice. Two curves cross at $J_x/J_z$=1.}
\label{fig:correlationfunction}
\end{figure}

\begin{figure}[h]
\includegraphics[width=9cm]{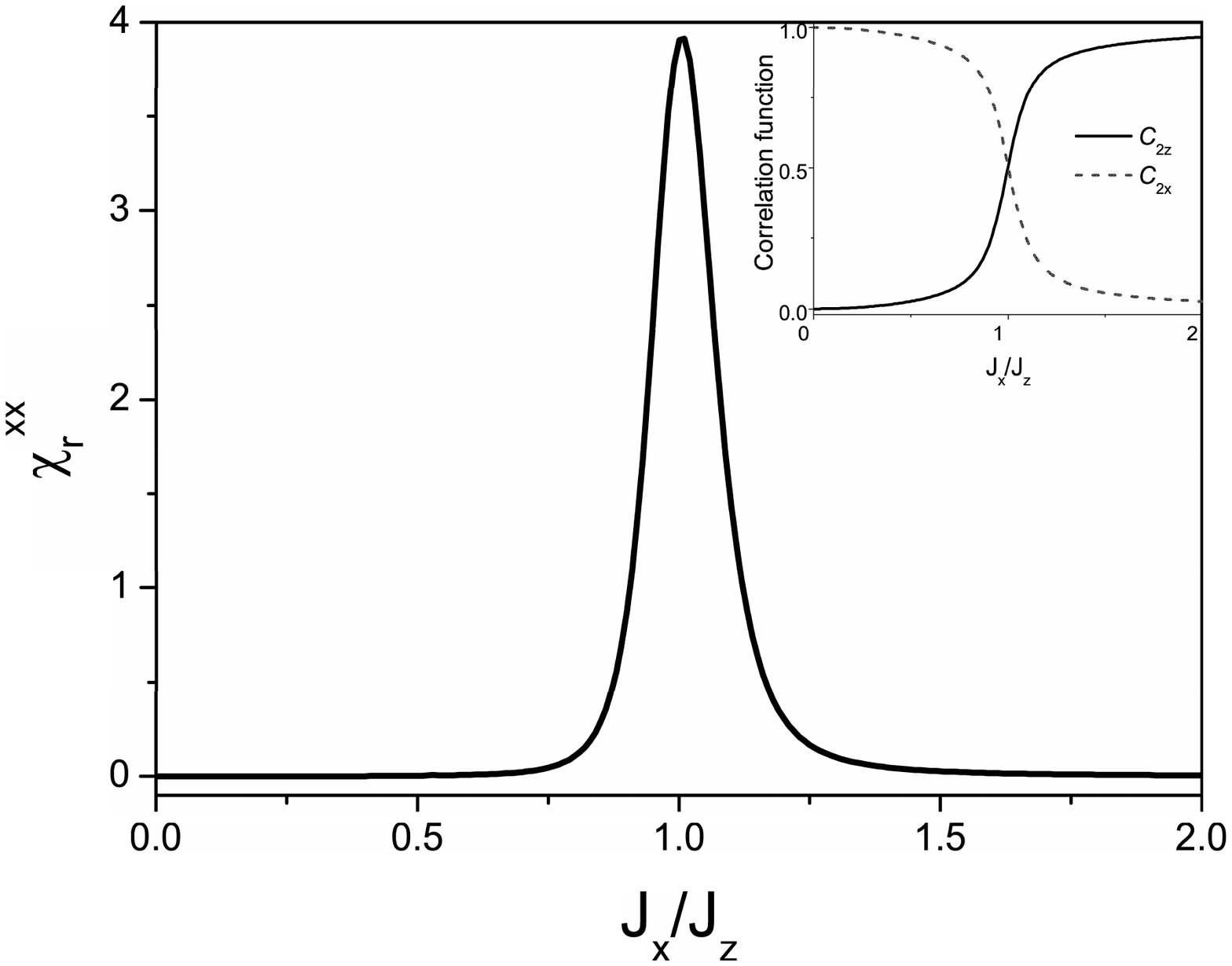}
\caption{(Color online) The reduced fidelity susceptibility of two
next-nearest-neighbor sites for $N=18$ square lattice with respect
to $J_x/J_z$. Inset shows the next-nearest-neighbor correlation
functions verse $J_x/J_z$. } \label{fig:NNNCF-N18}
\end{figure}

At this stage, a single-site pseudospin-flipping interaction will
change the parity of the chain, contrary to the flipping terms in
$H$. For example, $\hat{S}_{\bf i}^{x}$ or $\hat{S}_{\bf i}^{y}$
acting on one chain along $x$ direction will change the parity of
the chain. For two-point correlation functions between sites ${\bf
i}$ and ${\bf j}$ separated by $n$ sites, it is not difficult to
yield
\begin{eqnarray}
e_{nz}^x &=& (-1)^n \left\langle  \Psi_0(\Lambda)\left\vert
\hat{S}_{\bf i}^{x} \hat{S}_{{\bf i}+n \hat{z}}^{x}\right\vert
\Psi_0(\Lambda)\right\rangle =0.
\end{eqnarray}
Similarly, $e_{nz}^y$=0, $e_{nx}^y=0$, $e_{nx}^z=0$. The only
correlation functions surviving are $\mathcal{C}_{nx} \equiv 4
e_{nx}^x $ and $\mathcal{C}_{nz} \equiv 4 e_{nz}^z $, as are shown
in Fig. \ref{fig:correlationfunction} and inset of Fig.
\ref{fig:NNNCF-N18}. As $J_x$ increases, $\mathcal{C}_z$ decreases
while $\mathcal{C}_x$ increases accordingly. They become equal at
$J_x$= $J_z$.

With these properties of the ground state $\vert \Psi_0\rangle$ of
the 2D QCM, a simple form for the two-site density matrix is
obtained,
\begin{equation}
   \rho(\mathbf{i},\mathbf{j})=\mathbf{Tr}'(|\Psi_0\rangle\langle \Psi_0|)=\frac{1}{4}\sum_{\alpha,\alpha '=0}^3\langle \sigma_{\mathbf{i}}^{\alpha}\sigma_{\mathbf{j}}^{\alpha'}\rangle
   \sigma_{\mathbf{i}}^{\alpha}\sigma_{\mathbf{j}}^{\alpha'},
\end{equation}
in which the prime means tracing over all the other pseudospin
degrees of freedom except the two sites $\mathbf{i}$ and
$\mathbf{j}$. $\sigma^{\alpha}$ are Pauli matrices $\sigma^{x}$,
$\sigma^{y}$ and $\sigma^{z}$ for $\alpha$ = 1 to 3, and 2 by 2 unit
matrix for $\alpha$=0. If the two pseudospins are linked by an
$x$-type bond, we have
\begin{equation}
   \rho_x(\mathbf{i},\mathbf{j})=\mathbf{Tr}'(|\Psi_0\rangle\langle \Psi_0|)=4\langle S_{\mathbf{i}}^x S_{\mathbf{j}}^x\rangle S_{\mathbf{i}}^x
   S_{\mathbf{j}}^x+\frac{1}{4}I_{\mathbf{i}}I_{\mathbf{j}},
\end{equation}
in which $I_{\mathbf{i}}$ and $I_{\mathbf{j}}$ are the 2 by 2 unit
matrices. If translational invariance is preserved in the ground
state $|\Psi_0\rangle$, the reduced density matrix can be simplified
as
\begin{equation}
     \rho_x(\mathbf{i},\mathbf{j})=-\mathcal{C}_x
     S_{\mathbf{i}}^xS_{\mathbf{j}}^x+\frac{1}{4}I_{\mathbf{i}}I_{\mathbf{j}}.
     \label{rho_x}
\end{equation}
Similarly, if the two sites $\mathbf{i}$ and $\mathbf{j}$ are linked
by $z$-type bond, we have
\begin{equation}
\rho_z(\mathbf{i},\mathbf{j})=-\mathcal{C}_z
S_{\mathbf{i}}^zS_{\mathbf{j}}^z+\frac{1}{4}
I_{\mathbf{i}}I_{\mathbf{j}}.\label{rho_z}
\end{equation}

In addition, the NN two-point correlation functions
$\mathcal{C}_\alpha$ can be calculated using the Feymann-Hellmann
Theorem
\begin{equation}
\mathcal{C}_\alpha=-\frac{4}{N}\frac{\partial E_0}{\partial
J_\alpha}, \label{C_alpha}
\end{equation}
where $\alpha=x,z$ means $x$-type and $z$-type bond respectively.

\begin{figure}[h]
\includegraphics[width=9cm]{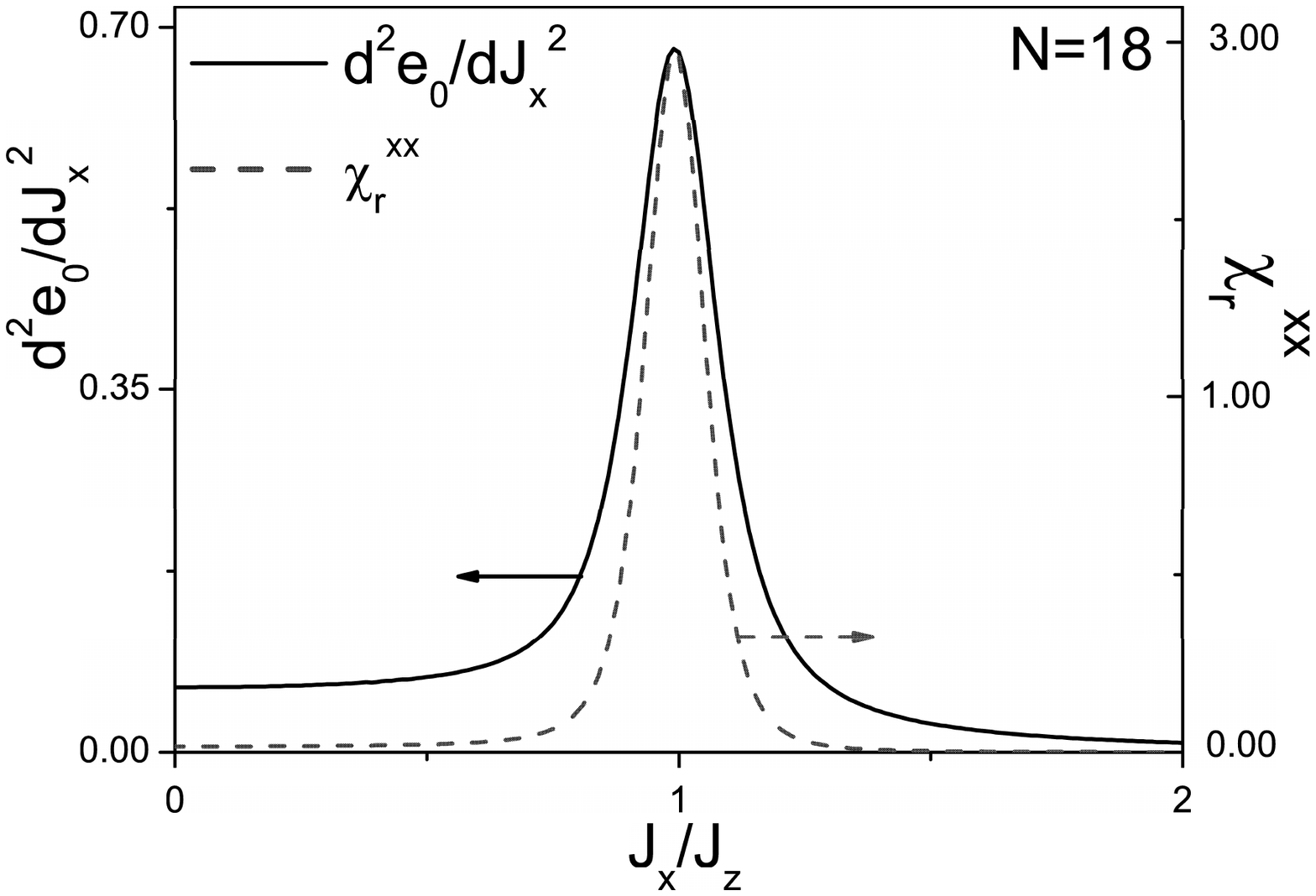}
\caption{(Color online) The second derivative of ground state energy
(solid line) and reduced fidelity susceptibility (dash line) of
$N=18$ square lattice as a function of $J_x/J_z$.}
\label{fig:RFS-E-18}
\end{figure}

The reduced fidelity $F_{\textrm{r}}$ is defined as the overlap
between $\rho_\alpha(\mathbf{i},\mathbf{j})$ and
$\rho'_\alpha(\mathbf{i},\mathbf{j})$, i.e.
\begin{equation}
  F_{\textrm{r}}(\rho_\alpha(\mathbf{i},\mathbf{j}),\rho'_\alpha(\mathbf{i},\mathbf{j}))=\mathbf{Tr}\sqrt{\rho_\alpha^{1/2}\rho_\alpha'\rho_\alpha^{1/2}}.
\end{equation}
The prime means a different reduced density matrix induced by the
tiny changes of driving parameters in the Hamiltonian. By Eqs.
(\ref{rho_x}) and (\ref{rho_z}), it is obvious that $\rho_\alpha$
commutes with $\rho'_\alpha$, and we can easily evaluate the reduced
fidelity
\begin{eqnarray}
   F_{\textrm{r}}(\rho_\alpha(\mathbf{i},\mathbf{j}),\rho'_\alpha(\mathbf{i},\mathbf{j}))&=&\frac{1}{2}\left(\sqrt{(1+\mathcal{C}_{\alpha})(1+\mathcal{C}'_{\alpha})} \right. \nonumber \\
   &+&
   \left.\sqrt{(1-\mathcal{C}_{\alpha})(1-\mathcal{C}'_{\alpha})}\right).
\end{eqnarray}
The two-site RFS $\chi_{\textrm{r}}$ could also be obtained
straightforwardly,
\begin{eqnarray}
  \chi_{\textrm{r}}^{\alpha \beta} &=& \lim_{\delta J_{\beta} \to 0}\frac{-2 \ln F_{\textrm{r}}}{(\delta J_{\beta})^2}
  =
  \frac{(\partial_{J_{\beta}}\mathcal{C}_{\alpha})^2}{4(1-\mathcal{C}_{\alpha}^2)},
  \label{DefinitionofRFS}
\end{eqnarray}
in which $J_{\beta}$ ($\beta$=$x$,$z$) is the driving parameter for
the QPT. According to Eqs. (\ref{C_alpha}) and
(\ref{DefinitionofRFS}), we observe that the numerator of
$\chi_{\textrm{r}}^{\alpha \alpha}$ is proportional to the square of
the second derivative of GS energy and the denominator is finite in
unpolarized state. The second power in the numerator indicates that
the two-site RFS is more effective than the second derivative of the
GS energy in measuring QPTs \cite{Xiong-PRB-2009}. We compare the
RFS of two NN sites with the second derivative of GS energy of
$N=18$ square lattice (see Fig. \ref{2Dsquarelattice}) displayed in
Fig. \ref{fig:RFS-E-18}, and find that they indeed present
cusp-shaped peaks, but the former is more sensitive to phase
transition than the latter. The pronounced maxima of peaks arise at
$J_x/J_z=1$.

\begin{figure}[h]
\includegraphics[width=9.5cm]{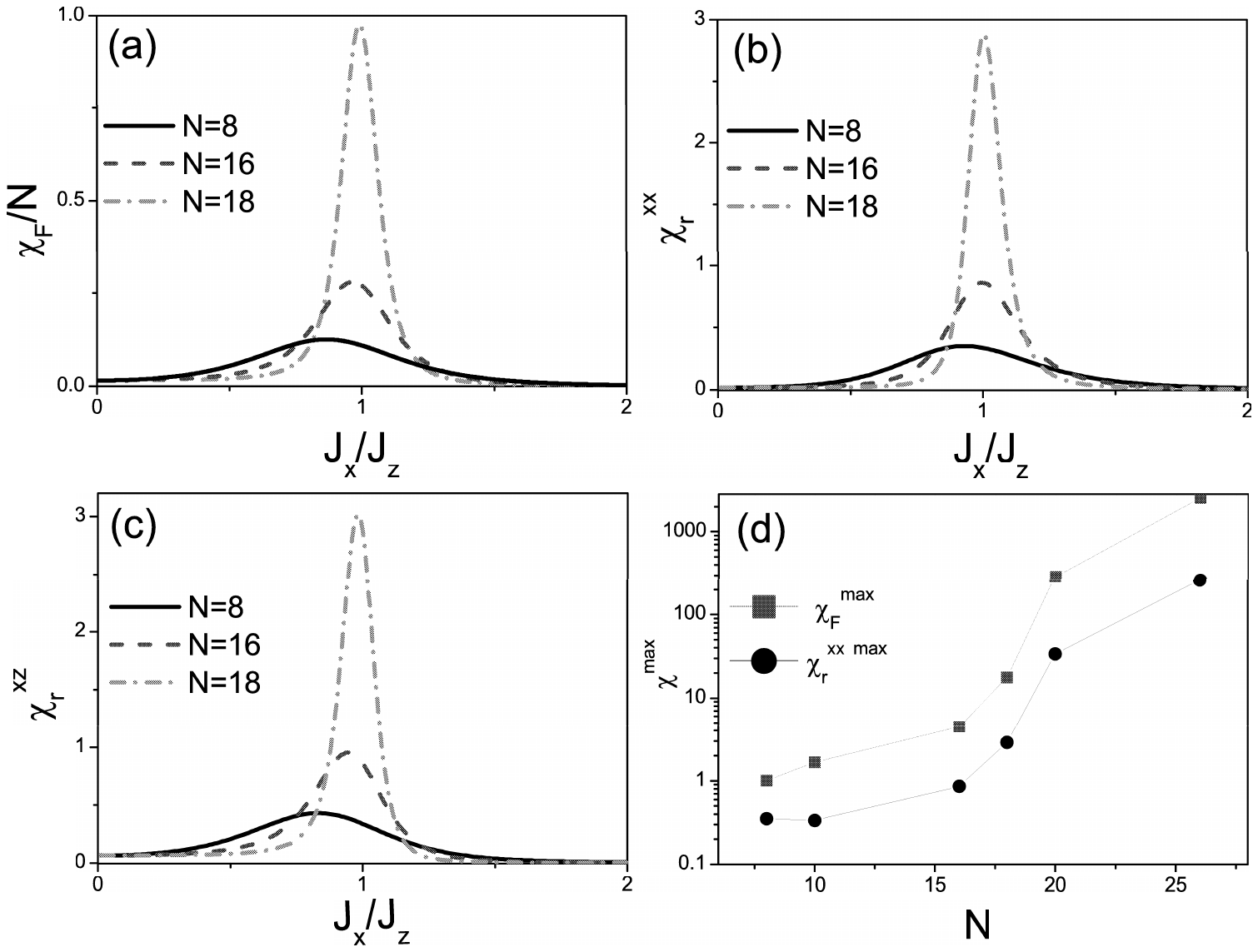}
\caption{(Color online) (a) The fidelity susceptibility and (b) the
reduced fidelity susceptibility of $x$-bond and (c) $z$-bond in the
ground state of the 2D compass model on a square lattice as a
function of $J_x/J_z$. The curves correspond to different lattice
sizes $N$=8, 16 and 18, respectively. (d) The log-linear plot of
maximum fidelity susceptibility and maximum reduced fidelity
susceptibility against the system size $N$.} \label{fig:XF-Jx-N}
\end{figure}

Fig. \ref{fig:XF-Jx-N} unveils the results from ED on various 2D
square lattices with PBC. As the coupling $J_x$ changes, both FS and
RFS of two NN sites exhibit peaks around $J_x =J_z$. With increasing
system size, the peaks of $\chi_{\textrm{F}}$ (or
$\chi_{\textrm{r}}$) become more pronounced and pseudocritical
points seemingly converge toward the real critical point $J_x =J_z$
quickly. We plot the maximum fidelity
$\chi_{\textrm{F}}^{\textrm{max}}$ and maximum reduced fidelity
susceptibility $\chi_{\textrm{r}}^{\textrm{max}}$ against the system
size $N$ in Fig. \ref{fig:XF-Jx-N}(d). The scaling of the peaks
reveals approximately exponential divergences at criticality instead
of power-law divergences, which indicates the occurrence of
non-second-order phase transition.

As is argued in Eqs. (\ref{gap}) - (\ref{scalingof1storder}), the
approximately exponential divergence of FS and RFS hints an
exponentially small gap at the critical point and thus a first-order
phase transition. For two next-nearest-neighbor (NNN) sites, we
calculate the RFS according to Eq. (\ref{DefinitionofRFS}). We
observe a similar behavior of the RFS with respect to $J_x/J_z$, and
find that the RFS of NNN pair is a bit larger than the RFS of NN
bond. This is illustrated in Fig. \ref{fig:NNNCF-N18}. In a word,
the two-site RFS can serve as a signature for the QPTs in the 2D
compass model.

\section{Summary and Discussion}\label{Summary}
Motivated by recent theoretical and experimental work on orbital
degree of freedom in Mott insulators, we have studied the QPTs in
various 2D spin-orbit models using FS and RFS. The numerical
analysis is performed on 2D clusters with the Lanczos algorithm. The
spin-orbit model hosts plentiful phases, including phase transitions
within and beyond the framework of Landau-Ginzburg paradigm. For the
2D XXZ model with DMI, with increasing driving parameters, the
system undergoes a second-order phase transition from AFM state
along $x$-$y$ plane to AFM state along $z$ direction. We compare FS
with second derivative of GS energy, and find both of them exhibit
similar peaks. The FSS demonstrates that FS per site should diverge
in the thermodynamic limit at pseudocritical point, and the
locations of extreme points approach QCPs accordingly. The power-law
divergence of FS at criticality indicates the quantum phase
transition is of second order and the critical exponent $\nu$ is
obtained. Analogously, as the exchange coupling changes, the ground
state of 2D Kitaev-Heisenberg model evolves from N\'{e}el AFM state
to stripy AFM state, and to Kitaev spin liquid. The QCPs could be
signaled by the peaks of both FS and second derivative of GS energy.
The nonlocal symmetries in 2D AFM QCM guarantee that the RFS can be
written in an analytical form as a function of the correlation
functions. Peaks of FS and two-site RFS take place around
$J_x$=$J_z$. The quasicritical point develops into critical point
quickly with increasing the system size. Scaling of the peaks
reveals an exponential divergence at criticality, which suggests
that a first-order phase transition happens. In conclusion, the FS
and RFS are effective tools in detecting diverse QPTs in 2D
spin-orbit models, and their scaling behaviors may hint the orders
of phase transitions.

\section{ACKNOWLEDGMENTS}
We acknowledge useful discussions with Wing-Chi Yu and J. Sirker.
Wen-Long You acknowledges the support of the Natural Science
Foundation of Jiangsu Province under Grant No. 10KJB140010 and the
National Natural Science Foundation of China under Grant No.
11004144. Yu-Li Dong acknowledges the support of the Specialized
Research Fund for the Doctoral Program of Higher Education (Grant
No. 20103201120002), Special Funds of the National Natural Science
Foundation of China (Grant No. 11047168) and the National Natural
Science Foundation of China (Grant No. 11074184).

\end{document}